\documentstyle[epsf,rotate,fleqn,espcrc2]{article}

\newcommand{\del}{\partial}

\newcommand{\rf}[4]{{\em {#1}} {\bf #2}, #3 (#4)}

\newcommand{\pr}{Phys.\ Rev.\ }

\newcommand{\beq}{\begin{equation}}
\newcommand{\eeq}{\end{equation}}
\newcommand{\beqa}{\begin{eqnarray}}
\newcommand{\eeqa}{\end{eqnarray}}
\newcommand{\Tr}{{\rm Tr}\,}
\newcommand{\muhat}{\hat{\mu}}

\newcommand{\oa}[1]{${\cal O}(a^{#1})$}
\newbox\rotbox

\title{Improved Landau Gauge Fixing and Discretisation Errors
\thanks{Supported in part by the Australian Research Council}}

\author{F.\ D.\ R.\ Bonnet\address{CSSM and The Department of Physics and 
	Mathematical Physics, University of Adelaide, Adelaide SA 5005, 
	Australia}, 
	\addtocounter{address}{-1} P.\ O.\ Bowman\addressmark, 
	\addtocounter{address}{-1} D.\ B.\ Leinweber\addressmark,
	D.\ G.\ Richards\address{Dept. of Physics and Astronomy, 
	University of Edinburgh, Edinburgh EH9 3JZ, Scotland \\ and
	TJNAF, 12000 Jefferson 
	Avenue, Newport News, VA 23606, USA.}
	and \addtocounter{address}{-2} A.\ G.\ Williams\addressmark
 	\addtocounter{address}{1}\address{Department of Physics and SCRI, 
	Florida State University, Tallahassee, Florida, USA}
	\thanks{Talk presented by AGW}}

\begin{document}

\begin{abstract} 
Lattice discretisation errors in the Landau gauge condition are
examined.  An improved gauge fixing algorithm in which ${\cal O}(a^2)$
errors are removed is presented.  ${\cal O}(a^2)$ improvement of the gauge 
fixing condition displays the secondary benefit of reducing the size of 
higher-order errors.
These results emphasise the importance of implementing an improved gauge
fixing condition.
\end{abstract}

\maketitle


\section{Introduction}
\label{sec:intro}

Gauge fixing in lattice gauge theory simulations is crucial for many
calculations e.g.\ the study of gauge dependent
quantities such as the gluon propagator~\cite{rapid_glu}.
However, the standard lattice Landau gauge condition~\cite{cthd} is the same 
as the continuum condition,
$\sum_\mu \del_\mu A_\mu = 0$, only to leading order in the lattice
spacing, $a$.  

The focus of this talk is to use
mean-field-improved perturbation theory~\cite{tadpole} to compare  
different lattice definitions of the Landau gauge, and quantify the sizes of
the discretisation errors. In particular, we derive a new
${\cal O}(a^2)$ improved Landau-gauge-fixing functional, and a method of
generalising this to \oa{n}.

\section{Lattice Landau Gauge}
\label{sec:formalism}

Gauge fixing on the lattice is achieved by maximising a functional whose
extremum implies the gauge fixing condition.  The usual Landau 
gauge fixing functional is \cite{cthd}
\begin{equation}
{\cal F}^{G}_{1}[\{U\}] = \sum_{\mu, x}\frac{1}{2} \mbox{Tr} \, \left\{
U^{G}_{\mu}(x) + U^{G}_{\mu}(x)^{\dagger} \right\},
\label{eqn:f1}
\end{equation}
where
$U^{G}_{\mu}(x) = G(x) U_{\mu}(x) G(x+\hat{\mu})^{\dagger}.$
By taking the functional derivative of (\ref{eqn:f1}), 
it can be shown that a maximum of that functional implies the continuum 
Landau gauge, with \oa{2} errors. 
It can further be shown that the gauge fixing condition
implies that
\beq
\sum_\mu \del_\mu A_\mu(x) 
= \sum_\mu \left\{ -\frac{a^2}{12} \del_\mu^3 A_\mu(x) 
- {\cal H}_1 \right\},
\label{eqn:err1}
\eeq
where ${\cal H}_1$ represents ${\cal O}(a^4)$ and higher-order terms.
Na\"{\i}vely one might hope that higher-order derivatives in the
brackets are small, but it will be shown
that the terms on the R.H.S. of (\ref{eqn:err1}) are
large compared to the numerical accuracy possible in gauge fixing
algorithms.


This ``one-link'' functional can be generalised to functionals using 
``n-link'' terms:
\beq
{\cal F}_n^G[{U}] \equiv \sum_{x,\mu} \frac{1}{2n^2} \Tr \left\{ 
	U_{n\mu}^G (x) + \mbox{h.c.} \right\}
\eeq
where \\
$ U_{n\mu}^G(x) \equiv U_\mu^G(x) U_\mu^G(x + \muhat)...U_\mu^G(x + (n-1)\muhat) $.
Then
\begin{eqnarray}
&& \hspace{-7mm}\frac{\delta{\cal F}_n^G}{\delta \omega^a(x)} 
	 = \frac{1}{2n^2}i \sum_\mu \Tr \left\{ \left [ 
		U_{n\mu}^G (x-\muhat) - U_{n\mu}^G(x) \right. \right.
		\nonumber \\ 
&& \hspace{7mm}    \left. \left. - \mbox{h.c.} \right ] T^a \right\} 
\label{eqn:delta_fn} \\
%
&& \hspace{-7mm} =  ga^2 \Biggl( \sum_\mu \Tr \Biggl\{ \biggl[ \del_\mu A_\mu (x) 
           + \frac{2}{(na)^2} \biggl\{ 
		\frac{{(na)}^4}{4!}
		\del_\mu^3 A_\mu (x)  
		\nonumber \\
&& \hspace{5mm}	 +  \frac{{(na)}^6}{6!}
		\del_\mu^5 A_\mu(x) + ... \biggr\} \biggr] T^a \Biggr\}
		+ {\cal O}(g^2a^2) \Biggr)  \nonumber
\end{eqnarray}

${\cal O}(a^2)$ errors can be removed from the gauge fixing condition
by taking a linear combination of the one-link and two-link
functionals:
\begin{equation}
{\cal F}_{\rm Imp}^G = \frac{4}{3} {\cal F}_1^G 
			- \frac{1}{12 u_0} {\cal F}_2^G
\label{eqn:fimp}
\end{equation}
where we have included the plaquette-based, mean-field (tadpole) improvement 
parameter, $u_0$~\cite{tadpole}.

To perform the gauge fixing we adopt a ``steepest descents'' 
approach~\cite{cthd}.  The gauge transformation is 
$G(x) = \exp\{ - i\alpha \sum_\mu \del_\mu A_\mu (x) \}$.
%
To maximise, for example, ${\cal F}_1^G$, we use (\ref{eqn:delta_fn}) 
to derive the gauge transformation
\beq
G_1(x) = \exp \left\{ \frac{\alpha}{2} \Delta_1(x) \right \},
\eeq
where
\begin{displaymath}
\Delta_1(x) = \sum_\mu \left\{ U_\mu^G (x-\muhat) - U_\mu^G (x) 
		 - \mbox{h.c.} \right\}_{\rm traceless}.
\end{displaymath}
Similarly, $\Delta_2$ and $\Delta_{\rm Imp}$ are obtained from the
functional derivatives of ${\cal F}_2$  and ${\cal F}_{\rm Imp}$ respectively. 
For a given functional, ${\cal F}_i^G$, the gauge fixing algorithm proceeds 
by calculating the relevant $\Delta_i$,
applying the associated gauge transformation to the gauge field, and 
iterating until the lattice Landau gauge condition
is satisfied, to within some numerical accuracy.
The approach to Landau gauge is measured by
\beq
\theta_i = \frac{1}{V N_c} \sum_x \mbox{Tr} \left\{ \Delta_i(x) 
\Delta_i(x)^{\dagger}
\right\}.
\eeq

A configuration fixed using $\Delta_1(x)$ will satisfy (\ref{eqn:err1}).
It will also satisfy
\begin{displaymath}
\Delta_2(x) = -2iga^2 \sum_\mu \left\{ \frac{a^2}{4} 
		\del_{\mu}^3 A_{\mu}(x) 
	 	- {\cal H}_1 + {\cal H}_2 \right\}
\end{displaymath}
and similarly,
\beqa
\Delta_{\rm Imp}(x) & = & -2iga^2 \sum_\mu \left\{ -\frac{a^2}{12} 
		\del_{\mu}^3 A_{\mu}(x) \right. \nonumber \\
	& & \left. - {\cal H}_1 + {\cal H}_{\rm Imp} \right\}.
\label{eqn:deltaImp_err}
\eeqa
Since the improved measure has no ${\cal O}(a^2)$ error of its own,
(\ref{eqn:deltaImp_err}) provides an estimate of the absolute size of 
these discretisation errors.

\section{Calculations on the Lattice}
\label{sec:results}

We use an ${\cal O}(a^2)$ tadpole-improved action.  
For the exploration of gauge fixing errors we consider
$6^4$ lattices at $\beta = 3.92$, 4.38, and 5.00, corresponding to lattice 
spacings of
approximately 0.35, 0.17, and 0.1 fm respectively.

\begin{figure}
\epsfysize=7.2truecm
\setbox\rotbox=\vbox{\epsfbox{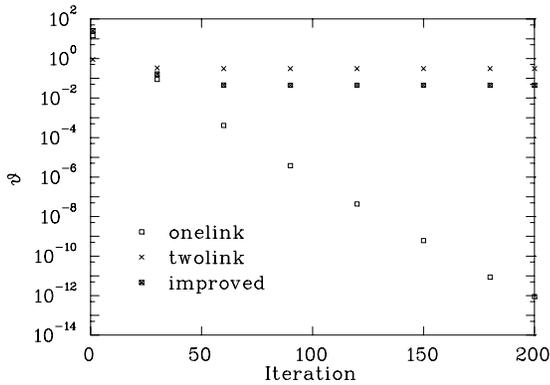}}\rotl\rotbox
\vspace{-12pt}
\caption{The gauge fixing measures for a $6^4$ lattice with Wilson action 
at $\beta = 6.0$.  This lattice was gauge fixed with $\Delta_1$, so 
$\theta_1$ drops steadily whilst $\theta_2$ and $\theta_{\rm{Imp}}$
plateau at much higher values.}
\label{fig:thetas}
\end{figure}

The configurations are gauge fixed, using Conjugate Gradient Fourier 
Acceleration~\cite{cm} until $\theta_1 < 10^{-12}$.  $\theta_{\rm{Imp}}$
and $\theta_2$ are then measured, to see the size of the residual higher
order terms.  The evolution of the gauge fixing measures is shown for one of
the lattices in Fig.~\ref{fig:thetas}.  This procedure is then repeated, 
fixing with each of the other two functionals, and
the results are shown in Table~\ref{tab:imp_s6t6}.  
Results from additional lattices, as well as a more detailed discussion, are
in~\cite{pob}.

\begin{table}
\caption{Values of the gauge-fixing measures obtained using the improved 
gluon action on $6^4$ lattices at three values of the lattice spacing, fixed to
Landau gauge with the one-link, two-link and improved functionals 
respectively.}
\begin{tabular}{ccclll}
\hline
$\beta$ & $u_0$ & $\cal F$ & $\theta_{\rm{Imp}}$ & $\theta_2$
& $\frac{\theta_{\rm{Imp}}}{\theta_2}$ \\ \hline
  3.92  & 0.837 & 1 & 0.102  & 0.921 & 0.111 \\
  4.38  & 0.880 & 1 & 0.0585 & 0.526 & 0.111 \\
  5.00  & 0.904 & 1 & 0.0410 & 0.369 & 0.111 \\
\hline
\hline
$\beta$ & $u_0$ & $\cal F$ & $\theta_{\rm{Imp}}$ & $\theta_1$ 
& $\frac{\theta_1}{\theta_{\rm{Imp}}}$ \\ 
\hline
  3.92  & 0.837 & 2 &  57.5  & 32.3 & 0.563 \\
  4.38  & 0.880 & 2 &  53.4  & 30.0 & 0.563 \\
  5.00  & 0.904 & 2 &  52.2  & 29.4 & 0.563 \\
\hline
\hline
$\beta$ & $u_0$ & $\cal F$ & $\theta_1$ & $\theta_2$ 
& $\frac{\theta_1}{\theta_2}$ \\ 
\hline
  3.92  & 0.837 & Imp & 0.0638 & 1.02  & 0.0625 \\
  4.38  & 0.880 & Imp & 0.0366 & 0.586 & 0.0625 \\
  5.00  & 0.904 & Imp & 0.0261 & 0.417 & 0.0625 \\
\hline
\end{tabular}
\label{tab:imp_s6t6}
\end{table}

If we fix a configuration to Landau gauge by using the basic, one-link
functional, then the improved measure, $\theta_{\rm Imp}$, will consist 
entirely of the discretisation errors.  Looking at Table~\ref{tab:imp_s6t6}, 
we see that at
$\beta = 4.38$, $\theta_{\rm Imp} = 0.059$,
a substantial deviation from the continuum Landau gauge compared to the
tolerance of the gauge fixing.
We note that the relationship between the functionals in (\ref{eqn:fimp})
provides a constraint on the gauge fixing measures.
For example, when fixing with $\Delta_1$
\beq
\frac{\theta_{\rm{Imp}}}{\theta_2}  =  \frac{(-\frac{1}{12})^2}
		{(-\frac{1}{12} + \frac{1}{3})^2}
		=  \frac{1}{9} \simeq  0.111.
\eeq
A configuration fixed using $\Delta_{\rm{Imp}}(x)$ will satisfy 
\beq
\sum_\mu \del_\mu A_\mu(x) 
= \sum_\mu \left\{ - {\cal H}_{\rm{Imp}} \right\}.
\label{eqn:errImp}
\eeq
With the help of (\ref{eqn:err1}) we see that
\begin{displaymath}
\Delta_1(x) = -2iga^2 \sum_\mu \left\{ 
		\frac{a^2}{12} \del_\mu^3 A_\mu(x)  
		+ {\cal H}_1 - {\cal H}_{\rm{Imp}} \right\}
\end{displaymath}
and (\ref{eqn:deltaImp_err}) are identical to within a sign.
If the three different methods presented all fixed in exactly the same way,
then the $\theta_{\rm{Imp}}$ of a configuration fixed with $\Delta_1$, would
be equal to $\theta_1$ when the configuration is fixed with 
$\Delta_{\rm{Imp}}$.  It is clear from the table that they are not, 
signaling the higher-order derivative terms $\del_\mu^n A_\mu(x)$, contained
in the the ${\cal H}_i$, take
different values depending on the gauge fixing functional used.

Examining the values in Table~\ref{tab:imp_s6t6} reveals that
in every case $\theta_1$ is smaller when we have fixed with 
${\cal F}_{\rm Imp}$ than $\theta_{\rm{Imp}}$ under the 
${\cal F}_1$.  This
suggests that the additional long range information used by the improved 
functional is producing a gauge fixed configuration with smaller, higher-order
derivatives; a secondary effect of improvement.

Equally, one can compare the value of $\theta_2$ when fixed using
${\cal F}_1$, and $\theta_1$ when fixed using the ${\cal F}_2$.  In this 
case, their differences are rather large and are once again attributed to 
differences in the size of higher-order derivatives of the gauge field.  
${\cal F}_2$ is coarser, knows little about short range fluctuations, and 
fails to constrain higher-order derivatives.  Similar conclusions are
drawn from a comparison of $\theta_2$ fixed with the ${\cal F}_{\rm Imp}$
and $\theta_{\rm{Imp}}$ fixed with ${\cal F}_2$.

We also find that in terms of the absolute errors,
the Wilson action at $\beta = 6.0$ is comparable to the improved
lattice at $\beta = 4.38$, where the lattice spacing is three times
larger.

\section{Conclusions}

We have fixed gluon field configurations to Landau gauge by three
different functionals: one-link and two-link functionals, both with
${\cal O}(a^2)$ errors, and an improved functional, with ${\cal
O}(a^4)$ errors.  Using these functionals we have devised a method for
estimating the discretisation errors involved.  Lattice Landau gauge, in its
standard implementation, deviates from its continuum
counterpart by one part in 20, despite fixing the Lattice gauge condition 
to one part in $10^{12}$.  Our results indicate
that order ${\cal O}(a^2)$ improvement of the gauge fixing condition
improves comparison with the continuum Landau gauge through:
1) the elimination of ${\cal O}(a^2)$ errors and 2) reducing the size of 
higher-order errors.



\begin{thebibliography}{99}

\bibitem{rapid_glu} D.B.~Leinweber {\it et al.}\/ 
	\rf{\pr}{D 58}{031501}{1998}, hep-lat/9803015;
 	{\it ibid.}, \rf{\pr}{D 60}{094507}{1999}, hep-lat/9811027,  
	and references therein. 

\bibitem{cthd} C.T.H.~Davies et al, \rf{\pr}{D 37}{1581}{1988}.

\bibitem{tadpole} G.P.~Lepage \& P.B.~Mackenzie, \rf{\pr}{D 48}{2250}{1993}.

\bibitem{cm} A.~Cucchieri \& T.~Mendes, \rf{\pr}{D 57}{3822}{1998}.

\bibitem{pob} F.D.R.~Bonnet {\it et al.}, {\it Austral.J.Phys.} {\bf 52} (1999) 939-948, hep-lat/9905006

\end{thebibliography}
\end{document}